\documentclass[a4paper,10pt,onecolumn,amsmath,amssymb,aps,prb]{revtex4}
\usepackage{graphicx,amssymb}
\usepackage{natbib}
\usepackage[left=0.95in]{geometry}
\usepackage{color}
\usepackage{graphicx}
\usepackage{subfigure}
\usepackage{dcolumn}
\usepackage{bm}

\usepackage{amsmath}
\begin{document}
\parskip=0.25pc
\parindent=2pc
\baselineskip=14pt
\newcommand\ds{\displaystyle}
\def\be{\begin{equation}}
\def\ee{\end{equation}}
\def\bea{\begin{eqnarray}}
\def\eea{\end{eqnarray}}
\def\del{\hbox{\bm\char "72}}
\def\Frac#1#2{{{\ds #1} \over {\ds #2}}}
\def\pder#1#2{{{\ds \partial #1} \over {\ds \partial #2}}}
\def\der#1#2{\Frac{{ d} #1}{{d} #2}}
\def\mb{\mbox}
\def\ul{\underline}
\def\wrt{with respect to}
\newcommand{\pbox}{\parbox[t]{6in}}

\title{Stability of viscosity stratified flows down an incline: Role of miscibility and wall slip}

\author{Sukhendu Ghosh\footnote{Email: sukhendu.math@gmail.com} and R.Usha\footnote{Email: ushar@iitm.ac.in}}
\affiliation{Department of Mathematics, Indian Institute of Technology  Madras, Chennai 600036, India.}
\date{\today}
\begin{abstract}
The effects of wall velocity slip on the linear stability of a gravity-driven miscible two-fluid flow down an incline are examined. The fluids have the matched density but different viscosity. A smooth viscosity stratification is achieved due to the presence of a thin mixed layer between the fluids. The results show that the presence of slip exhibits a promise for stabilizing the miscible flow system by raising the critical Reynolds number at the onset and decreasing the bandwidth of unstable wave numbers beyond the threshold of the dominant instability. This is different from its role in the case of a single fluid down a slippery substrate where slip destabilizes the flow system at the onset. Though the stability properties are analogous to the same flow system down a rigid substrate, slip is shown to delay the surface mode instability for any viscosity contrast. It has a damping/promoting effect on the overlap modes (which exist due to the overlap of critical layer of dominant disturbance with the mixed layer) when the mixed layer is away/close from/to the slippery inclined wall. The trend of slip effect is influenced by the location of the mixed layer, the location of more viscous fluid and the mass diffusivity of the two fluids. The stabilizing characteristics of slip can be favourably used to suppress the non-linear breakdown which may happen due to the coexistence of the unstable modes in a flow over a substrate with no slip. 
The results of the present study suggest that it is desirable to design a slippery surface with appropriate slip sensitivity in order to meet a particular need for a specific application.   
\end{abstract}
\maketitle

\section{Introduction}
The stability characteristics of a gravity-driven free surface flow over a slippery inclined plane examined by Pascal \cite{Pascal99a} within the framework of Orr-Sommerfeld analysis shows that the effects of wall slip on the primary instability is nontrivial. This is further confirmed by the long-wave theory and boundary layer approximation employed by Sadiq \& Usha \cite{Sadiq08a} and Samanta $et$ $al.$ \cite{Samanta11a}. Velocity slip at the wall destabilizes the flow system by lowering the critical Reynolds number and it stabilizes the short waves beyond the threshold for instability. Samanta $et$ $al.$. \cite{Samanta11a}, present the mechanism of the primary instability using Whitham wave hierarchy. The kinematics waves propagate much faster than the dynamic waves resulting in the instability of the flow system. Slip at the wall decelerates the dynamic waves; stabilizes the base flow by decelerating the dynamic waves, thereby contributing to enhancement of instability. Far from instability threshold, at large Reynolds numbers, the base flow accelerates in the presence of slip; the film thickness decreases and the surface tension damping becomes more effective and this results in stabilizing effect of slip beyond the threshold for instability.

These studies which deal with slip effects are very relevant and significant since, in several natural and industrial settings, the solid substrates are permeable. Beavers and Joseph \cite{Beavers67} proposed a semi-empirical velocity slip boundary condition while examining the macroscopic model of transport phenomena across a fluid and a porous medium interface. The condition accounts for the local geometry of the interface through a dimensionless slip coefficient. It is clear from the investigations by Blake, Vinogradova and Vorono $et$ $al.$ \cite{Blake90,Vinogradova95,Vorono08} that imposing a slip boundary condition at a smooth solid-liquid interface takes into account the roughness of the substrate at a microscale, superhydrophobic surfaces and grooved surfaces.

Miksis \& Davis \cite{Miksis94} have derived an effective boundary condition for a film over a rough surface and in the case of a single-phase flow, the condition is a Navier-slip condition with slip-coefficient equal to the average amplitude of the roughness, if the amplitude of the roughness is small. Min and Kim \cite{Min05a} have considered the effects of hydrophobic surfaces on flow stability and transition to turbulence in view of its relevance in many engineering applications. In their study, the hydrophobic surface is represented by a slip boundary condition on the surface. Their results reveal that slip on the surface has significant impact on the stability characteristics and transition.

The slip condition can also be modeled by eddies over wavy/rough substrates \cite{Wierschem03,Wierschem04}
The experimental study by Wierschem $et$ $al.$ \cite{Wierschem03} on vortices in film flow over strongly undulated bottom substrates at small Reynolds numbers and authors have pointed out the creation and evolution of eddies in low Reynolds number flows beyond certain critical values of corrugation steepness and film thickness. The effects of inertia on the eddies formed in creeping films over strongly undulated substrates have been examined by Wierschem and Aksel \cite{Wierschem04}. It is important to note that the no-slip boundary condition does not hold at the seperatrix and the eddies act like fluids roller bearings \cite{Scholle08}

In addition, there are several settings and applications such as lubrication, microfluidics \cite{Zhu01a,Thompson97a}, polymer melt \cite{Denn01a}, where the velocity of a viscous fluid exhibits a tangential slip on the wall. In fact, slip effects have been investigated in a plane Poiseuille flow of a single fluid with both symmetric and asymmetric slip conditions \cite{Gersting74a,Lauga05a},
The results of linear stability analysis by Lauga \& Cossu \cite{Lauga05a} show that slip increases the critical Reynolds number for instability. The wall slip has a destabilizing role in flow through a diverging channel at low Knudsen numbers has been shown by Sahu $et$ $al.$ \cite{Sahu08a}. The stability of interface dominated immiscible two-fluids separated by a sharp interface \cite{You09a} shows that the boundary slip enhances the stability of the stratified microchannel flow.

In a recent article by Ghosh $et$ $al.$ \cite{Sukhendu14a}, the linear stability characteristics of pressure-driven miscible three-layer two-fluid slippery channel flow with matched density and varying viscosity has been examined. An interesting feature of the instability is that, at any Reynolds number, shorter wave lengths and smaller wave numbers are stable. The slip at the wall either stabilizes or destabilizes this flow system as compared to that in a rigid channel \cite{govindarajan04a}. The dual role exhibited by slip at the wall suggests that it is possible to control the stability properties of miscible two-fluid flow system with stratified viscosity in a channel by designing the walls of the channel as hydrophobic or rough or porous substrates with small permeability, which can be modeled by velocity slip at the walls.

An immediate curiosity is to examine the stability properties of a viscosity stratified two-layer free surface flow down an incline and analyze the effects of miscibility and wall slip on the flow system. Such an investigation gains its importance due to the necessity for understanding the effects of viscosity stratification in gravity-driven free surface flows down inclined porous/rough/hydrophobic substrate which can be modeled by substrate with velocity slip at the wall. 

In fact, there are numerous studies which are motivated by the need to improve the performance of many industrial processes, and which have focused their attention in understanding the effects of viscosity stratification in a Poiseuille flow through channels/pipes and in gravity-driven free surface flows down rigid inclined substrates
\cite{Govindarajan14a,Usha13a,Yih67,Hooper83,Hinch84,Renardy87a,Renardy92a,south99a,yiantsios88a,Kao65a,Kao65b,
Kao68,Wang78,Loewenherz89,Chen93,Jiang05,ranganathan01a,govindarajan04a,Malik05a,ern03a,Selvam07,Sukhendu14a,Craik69,Craik68,
Drazin62,Goussis87,sahu09a,talon11a,Lees46,pinarbasi95,laure97a,Weinstein90,Nouar07}.  
In these studies, viscosity stratification, that arises either due to $(i)$ a discontinuity in viscosity across a sharp interface of two immiscible fluids in contact \cite{Yih67,Hooper83,Hinch84,Renardy87a,Renardy92a,south99a,yiantsios88a,Kao65a,Kao65b,
Kao68,Wang78,Loewenherz89,Chen93,Jiang05},
or $(ii)$ by varying continuously the concentration or temperature across a film in which a diffusive interface of non-zero thickness occurs
\cite{ranganathan01a,govindarajan04a,Malik05a,ern03a,Selvam07,Sukhendu14a,Craik69,Craik68,
Drazin62,Goussis87,sahu09a,talon11a},
or $(iii)$ by using a Non-Newtonian fluid system \cite{pinarbasi95,laure97a,Weinstein90,Nouar07,Usha11}
is considered. In what follows, we focus on the investigations relevant to the present study, namely, those on stability characteristics of flows with viscosity stratification belonging to class $(ii)$ above.

The pioneering study by Craik \cite{Craik69} on the planar Couette flow in the long-wave limit clearly shows the important role played by the critical layer in the case of miscible systems. The results reveal that for a continuously stratified viscosity profile, viscosity stratification induces a curvature in the velocity profile and that if it is negative at the critical layer, then, it promotes instability. The stability characteristics of such a flow system is modified significantly and is different from those of flows with viscosity jumps. There is stabilization of the flow system when the viscosity decreases away from the interface (with critical layer of the flow in less viscous fluid); a reverse scenario is observed with a more viscous fluid adjacent to the wall. Craik \cite{Craik69} points out that the basic velocity has no point of inflection due to the change brought out by viscosity stratification and that the disturbances can not be neutral even in the absence of inertia.

The analysis by Craik and Smith \cite{Craik68} on the stability of free surface flows with viscosity stratification presents the long-wave limit results for arbitrary continuous viscosity distributions in a film over a rigid substrate. Their analysis for a viscosity distribution which is exponentially increasing with depth reveals that the surface mode ($S$ mode) that exists for thin films with viscosity increasing away from the interface is more stable than that with uniform viscosity. A finite wavelength analysis considered by Goussis and Kelly \cite{Goussis87} on the stability of a liquid film down a heated or cooled inclined substrate displays a similar effect with wall cooling. 
The study by Drazin \cite{Drazin62} on the stability of parallel flow of an incompressible fluid with variable density and viscosity incorporates the transport of viscosity by the mean flow but neglects the diffusion of viscosity in the transport equation for viscosity. This is similar to neglecting thermal diffusion in the temperature equation.

Lees and Lin \cite{Lees46} have pointed out that if diffusion is neglected, then the thermal equation for convection becomes singular; and hence one must account for the influence of thermal diffusivity in the neighbourhood of critical layers. This indicates that diffusivity between two superposed fluids must be accounted for while analysing the stability characteristics of miscible two-fluid flows.

The miscible two-fluid flow in which the layers of fluids are separated by mixed layer of finite-thickness belongs to the class $(ii)$ of viscosity stratified flows and has been analyzed by accounting for diffusion of two layers \cite{ranganathan01a,govindarajan04a,Malik05a,ern03a,sahu09a}. The effects of diffusion and thickness of mixed layer have been examined in detail by Ern et al. \cite{ern03a} for miscible two-fluid Couette flow. The growth rate is a  non-monotonic function of diffusion parameter and when the mixed layer thickness is not large, flows at intermediate Peclet numbers are unstable than those without diffusion. Further, viscosity stratified flows with thin mixed layer are shown to exhibit faster growing instability than flows with either sharp viscosity jump or continuous viscosity stratification across the entire flow. A new unstable mode, called the `$O$' mode is shown to exist due to the overlap of mixed layer with the critical layer of the dominant shear mode and this mode is absent in immiscible two-fluid flow. Govindarajan and co-workers \cite{ranganathan01a,govindarajan04a} have shown that the stability characteristics are influenced by viscosity ratio and that the flow becomes unstable at Reynolds number much lower than that for the corresponding immiscible configuration. 

The linear stability of miscible viscosity-stratified plane Poiseuille flow in Stokes' flow regime analyzed by Talon and Meiberg \cite{talon11a} reveals that $(a)$ diffusion destabilizes the flow system and $(b)$ short waves are destabilized when a highly viscous layer is in the core of the pipe. 

There are also applications, such as coating of a substrate or manufacture of photographic films and environmental flows such as rock glaciers in which one comes across instabilities in one and multiple layer free surface flows on inclined planes. In general, flow properties and fluids vary continuously in  a thin layer due to miscible nature of fluids. Usha $et$ $al.$ \cite{Usha13a} have analyzed this class of free surface flows down an inclined substrate where the flow contains a thin mixed layer between two fluids of different viscosity. Their results show that this class of miscible two-fluid viscosity stratified flow has stability characteristics that are qualitatively different from those of immiscible fluids and infinitely miscible fluids (having finite viscosity gradient throughout the film). The instabilities in this flow system arise due to the presence of a free surface and due to the interactions between critical layer of the viscosity transport and the mixed layer where there is viscosity gradient.

The above investigations reveal the effects of viscosity stratification in laminar channel flows with rigid walls and free surface flows over inclined rigid substrates with no-slip boundary condition. A series of wind-tunnel experiments performed in a channel flow by Sirovich \& Karlsson \cite{Sirovich97} reveal that by introducing specified patterns of protrusion on the confining walls of the channel, it is possible to achieve turbulence drag reduction. They have also reported that the arrangement of these protrusions resulted in either drag decrease or increased mixing. The results thus suggest a passive means of effectively controlling turbulence in channel flow.

The paper is organized as follows: After an overview of problem statement and methodology in Sec. II; the base flow computations, the validation results, and details of the pertinent stability properties are presented in Sec. III. 
Concluding remarks are presented in Sec. IV.

\section{Mathematical Formulation}
\label{sec:for}

\begin{figure}[!ht]
\centering
\includegraphics [scale = 0.35]{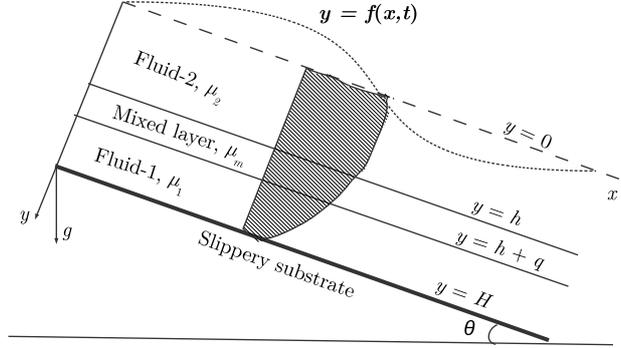}
\vspace{-1.5cm} 
\caption{Schematic of the flow system considered. Fluids `1' and `2' take the regions near the slippery inclined plane and near the free surface respectively. Both the fluids are separated by a mixed layer of uniform thickness $q$. The inclined slippery wall is located at $y = H$. $\theta$ is the angle of inclination.}
\label{fig1}
\end{figure}        

The linear stability of a two-dimensional, gravity-driven, miscible two-fluid flow down a slippery inclined substrate is considered. The fluids in layer $i$, $i = 1,2$ (Fig. \ref{fig1}) are density-matched ($\rho$), incompressible and Newtonian fluids with different viscosities $\mu_i$, $i = 1,2$. They are separated by a mixed layer ($h \leq y \leq h+q$) of viscosity stratified fluid of thickness $q$. The substrate is inclined at an angle $\theta$ to the horizontal. A cartesian co-ordinate system is chosen with origin at the unperturbed free surface, the $x$-axis along the unperturbed free surface ($y = 0$) parallel to the slippery inclined plane $y = H$. The $y$-axis is perpendicular to $y = 0$ and points towards the slippery wall. The free surface deflection is given by $y = f(x,t)$. The fluid layer-2 is adjacent to the free surface and is located in $0 \leq y \leq h$ and the layer-1 occupies the region $h+q \leq y \leq H$, near the slippery wall.

The governing equations of continuity, momentum transport, scalar-transport for viscosity and the boundary conditions at the slippery inclined wall and at the free surface are non-dimensionalized using the following scales: 

\begin{eqnarray}
x^{*} = \frac{x}{H},\, y^{*} = \frac{y}{H},\, t^{*} = \frac{U}{H}t, \,(u^{*},v^{*}) = \frac{1}{U}(u,v),\, p^{*} = \frac{1}{\rho  U^{2}}p,
\, \mu^{*} = \frac{\mu}{\mu_1}, \nonumber\\
h^{*} = \frac{h}{H},\,q^{*} = \frac{q}{H},\, m = \frac{\mu_2}{\mu_1},\, \beta = \frac{l_s}{H},\,\mu_m^*= {\frac{\mu_m(y)}{\mu_1}}. 
\label{ndv}
\end{eqnarray}

Here, $U$ corresponds to the average velocity across the film; $l_s$ is the slip length, $m$ is the viscosity ratio. $u,\, v$ are the velocity components in the $x$ and $y$ directions respectively; $p$ is the pressure and $t$ is the time. The dimensionless equations and boundary conditions are (after suppressing asterisks *),
\begin{equation}
{u_{x}}+{v_{y}} = 0 , \label{nd1}
\end{equation}
\begin{equation}
{u_{t}}+ {u}{u_{x}}+{v}{u_{y}} = \frac{\partial}{\partial x} \left [ - p + \frac{2}{Re}\mu u_x \right]
+ \frac{\partial}{\partial y} \left [ \frac{1}{Re}\mu (u_y + v_x) \right] + G , \label{nd2}
\end{equation}
\begin{equation}
{v_{t}}+ {u}{v_{x}}+{v}{v_{y}} = \frac{\partial}{\partial x} \left [ \frac{1}{Re}\mu (u_y + v_x) \right]
+ \frac{\partial}{\partial y} \left [ - p + \frac{2}{Re}\mu v_y \right] + G\cot{\theta} , \label{nd3}
\end{equation}
\begin{equation}
{\mu_{t}}+ {u}{\mu_{x}}+{v}{\mu_{y}} = \frac{1}{Pe} [\mu_{xx} + \mu_{yy}]. \label{nd4}
\end{equation}
At the free surface $y = f(x,t)$,
\begin{equation}
Re\,p = \frac{2\mu}{1+{f_x}^2}\left[ v_y - v_x f_x + u_x f_x^2 - u_y f_x \right] + \frac{S f_{xx}}{(1+f_x^2)^{3/2}}, \label{nd5}
\end{equation}
\begin{equation}
(1-f_x^2)(u_y + v_x) - 4u_xf_x = 0,  \label{nd6}
\end{equation}
\begin{equation}
v = f_t + uf_x. \label{nd7}
\end{equation}
At the slippery inclined plane,
\begin{equation}
u = - \beta {u_y}, \quad {v} = 0  \quad  \text {at} \quad y = 1. \label{nd8}
\end{equation} 
where $Re = \frac{\rho U H}{\mu_1}$, $Pe = \frac{U H}{\chi}$ ($\chi$ is the mass diffusivity) and $Sc = Pe/Re$ are the Reynolds, the Peclet and the Schmidt number respectively. $G = \frac{\rho g H^2 \sin{\theta}}{\mu_1 U}$ is the dimensionless gravity parameter, $S = \frac{\sigma}{\mu_1 U}$ is the dimensionless surface tension parameter. $g, \sigma$ are the gravitational force and the surface tension coefficient between the fluid and air at the free surface. $\beta$ is the slip parameter. 

In what follows, a linear stability of the base flow $U_B(y),\,P_B(y),\, \mu_B(y)$ that corresponds to a locally parallel flow of two fluids down a slippery inclined plane with a mixed layer of uniform thickness $q$ \cite{Usha13a} separating the two layers with constant viscosity $\mu_i$ ($i = 1,2$), is considered. The base velocity $U_B(y)$ and pressure $P_B(y)$ satisfy the following equations:
\begin{equation}
\frac{d}{dy}\left[ \mu_B(y) \frac{dU_B}{dy}\right] + G = 0,
\label{bs1}
\end{equation} 
\begin{equation}
\frac{dP_B}{dy} = \frac{G}{Re}\cot{\theta},
\label{bs2}
\end{equation}
where the base viscosity $\mu_B(y)$ is given by \cite{Usha13a}, 
\begin{equation}
\mu_B(y) = \left \{
{\begin{array}{lcl}
   m     ~~~~~~\quad {\rm if} \quad 0 \leq y \leq h, \\
   \mu_m(y)    \quad {\rm if} \quad h \leq y \leq h + q, \\
   1     ~~~~~~~\quad {\rm if} \quad h + q \leq y \leq 1.
\end{array}}\right.
\label{bs3}
\end{equation}
In the above equation $m = \mu_2/ \mu_1$ and the base viscosity $\mu_m(y)$ in the mixed layer is given by 
\begin{equation}
\mu_m(y) = \tanh(Ay+B),
\label{bs4}
\end{equation}       
where 
\begin{equation}
A = \frac{1}{q}\left[ \tanh^{-1}(1) -  \tanh^{-1}(m)\right], \nonumber
\end{equation} 
\begin{equation}
B = \tanh^{-1}(m)-\frac{h}{q}\left[ \tanh^{-1}(1) -  \tanh^{-1}(m)\right]. \nonumber
\end{equation}

The temporal stability characteristics of the base flow are examined using a linear stability analysis by considering two-dimensional disturbances of the flow variables. Using the normal mode form for the perturbations given by 
\begin{equation}
(u,v,p,s) = (U_B(y), 0, P_B(x), \mu_B(y)) + (\hat{u}, \hat{v}, \hat{p}, \hat{\mu})(y) \mathrm{e}^{\mathrm{i}\alpha(x - c t)},
\label{ptb}
\end{equation}
the following modified Orr-Sommerfeld \cite{Usha13a,drazinandreid} system is derived (after suppressing hat $(\;\hat{}\;)$)
\begin{eqnarray}
{\mathrm{i} \alpha \,Re} \left[ \phi {''}(U_B - c) - \alpha^2 \phi (U_B - c) - U_B{''} \phi \right] = {\mu_B \phi{''''}}  + 2\mu_B{'} \phi{'''}\nonumber \\
+ (\mu_B{''} - 2  \alpha^2 \mu_B)\phi{''} -2 \alpha^2 \mu_B{'} \phi{'} + (\alpha^2 \mu_B{''} + \alpha^4 \mu_B)\phi \nonumber\\
+ U_B{'} \mu{''} + 2U_B{''} \mu{'} + (U_B{'''} + \alpha^2 U_B{'})\mu,
\label{stab1}
\end{eqnarray}
\begin{equation}
{\mathrm{i} \alpha \, Pe} \left[ (U_B - c)\mu - \mu_B{'}\phi \right] = (\mu{''} - \alpha^2 \mu),
\label{stab2}
\end{equation}
\begin{equation}
\phi{'} = -\, \beta \phi{''}, \quad \phi = \mu = 0 \quad {\rm at} \quad y=1,
\label{bc1}
\end{equation}
\begin{equation}
\phi{''} + \alpha^2 \phi + U_B{''} \eta = 0  \quad {\rm at} \quad y=0,
\label{bc2}
\end{equation}
\begin{eqnarray}
\alpha \,Re (U_B - c)\phi{'} + \mathrm{i} \mu_B (\phi{'''} - 3\alpha^2 \phi{'}) +
\mathrm{i} \mu_B{'} (\phi{''} + \alpha^2 \phi) + 2 \mathrm{i} U_B{''} \mu\nonumber \\
~~~~~~ - \alpha(G cot\theta + \alpha^2 S )\eta = 0 \quad {\rm at} \quad y=0.
\label{bc3}
\end{eqnarray}
\begin{equation}
\phi + (U_B - c)\eta = 0 \quad {\rm at} \quad y=0,
\label{bc4}
\end{equation}
where prime $(')$ denotes differentiation with respect to $y$; $ \, \mathrm{i}\equiv \sqrt{-1}$; $\phi$, $\mu$ and $\eta$ are respectively the amplitudes of the disturbances of the stream function, viscosity and free surface. The flow is linearly unstable to the infinitesimal disturbance if $Im(c) > 0$, where $c$ is the wave speed and $\alpha$ (real and positive) is the wave number. In the absence of slip, the above system reduces to those by Usha $et$ $al.$ \cite{Usha13a} and when $Pe$ is set to infinity, to those by Craik and Smith \cite{Craik68}.
 
\section{Results and discussion}
\label{result}
A Chebyshev spectral collocation approach is employed to solve the eigenvalue problem governed by the system $(\ref{stab1}) - (\ref{bc4})$. The numerical solution is computed by modifying the stability code developed by Rama Govindarajan \cite{govindarajan04a} and used in miscible two-fluid flow down an incline by Usha $et$ $al.$ \cite{Usha13a}. In the film, along the $y$ axis, $n$ stretched collocation points $y_1,y_2,......,y_n$ defined by 
\begin{equation}
y_j = \dfrac{a}{\sinh(b y_0)}\,[\sinh\{(y_{c,j} - y_0)b\} + \sinh(b y_0)],
\end{equation}
are considered. Here, $a$ is the mid point of the mixed layer, $b$ represents the degree of clustering of the grid points around $y = a$ and
$y_{c,j}$ with $j = 1,2,...,n$ are Chebyshev collocation points, given by
\begin{equation}
y_{c,j} = 0.5 {\left\{\cos \left[ \pi \dfrac{(j - 1)}{(n - 1)} \right] + 1 \right\}},
\end{equation}
and
\begin{equation}
y_0 = \frac{1}{2b} \ln{\left[ \dfrac{1 + (\mathrm{e}^b - 1)a}{1 + (\mathrm{e}^{-b} - 1)a}\right]}.
\end{equation} 
Suitable grid stretching is achieved \cite{govindarajan04a,Usha13a} in the computations by taking the stretching parameter $b$ as $b = 8$. The eigenvalues are obtained using the open source software, LAPACK, and it is observed that for the range of parameters considered, the eigenvalues remained the same, when the number of collocation points was doubled from $n = 81$ to $n = 161$. This ensured the grid independency of the present computed results.

\begin{figure}[!ht]
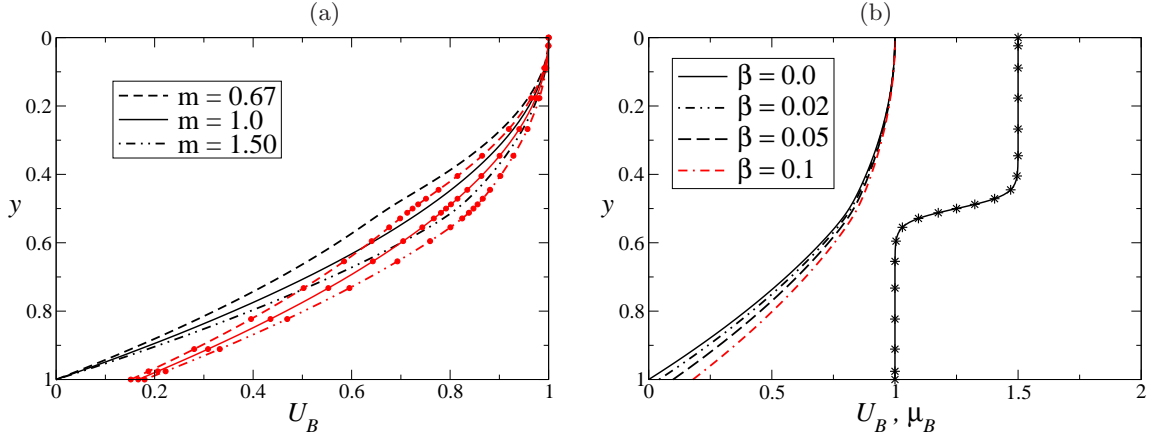

\centering
\hspace{0.2in} (a) \hspace{2.6in} (b) \\
\includegraphics [scale = 0.3]{fig2a.eps}\hspace{0.5cm}
\includegraphics [scale = 0.3]{fig2b.eps}\\
\caption{Base state velocity $U_B$ and viscosity $\mu_B$ (with star) profiles for $q = 0.2,\, h = 0.4$: (a) effect of viscosity ratio $m$ on $U_B$ and (b) Effect of the slip parameter $\beta$ on $U_B$ for $m = 1.5$. In Fig. (a), curves without/with symbols correspond to $\beta = 0$ and $\beta = 0.1$ respectively.}
\label{fig2}
\end{figure} 

Fig. \ref{fig2}(a) shows the base state velocity profiles for different viscosity stratifications when $q = 0.2, \theta = 10^{\circ}$ and $h = 0.4$. The base velocity $U_B(y)$ is a monotonic decreasing function of $y$ for any viscosity contrast ($m$). At a fixed distance from the unperturbed free surface ($y = 0$), $U_B(y)$ is more for a film over a slippery substrate ($\beta \neq 0$) as compared to that for $\beta = 0$ (Fig. \ref{fig2}(b)). Further, the wall stress decreases as slip parameter $\beta$ increases (Fig. \ref{fig2}(a),(b)). For the lubrication case corresponding to less viscous fluid adjacent to the inclined plane ($m = 1.5$), the velocity in the mixed layer ($0.4\leq y \leq 0.6$) is more and the profiles for all $\beta$ values considered are convex in nature (Fig. \ref{fig2}(b)). It is well known from inviscid theory that the flow system with base velocity more convex are inviscidly stable. Fig. \ref{fig2}(b) shows that when $m>1$, an increase in slip parameter increases the wall velocity. The base viscosity in this case for any $\beta$ is monotonically decreasing with respect to $y$ and at any $y$, it is independent of $\beta$. However, for the anti-lubrication case with more viscous fluid adjacent to the inclined plane ($m = 0.67$), the base viscosity, for any $\beta$, increases monotonically with an increase in $y$ (Fig. not shown).

\begin{figure}[!ht]
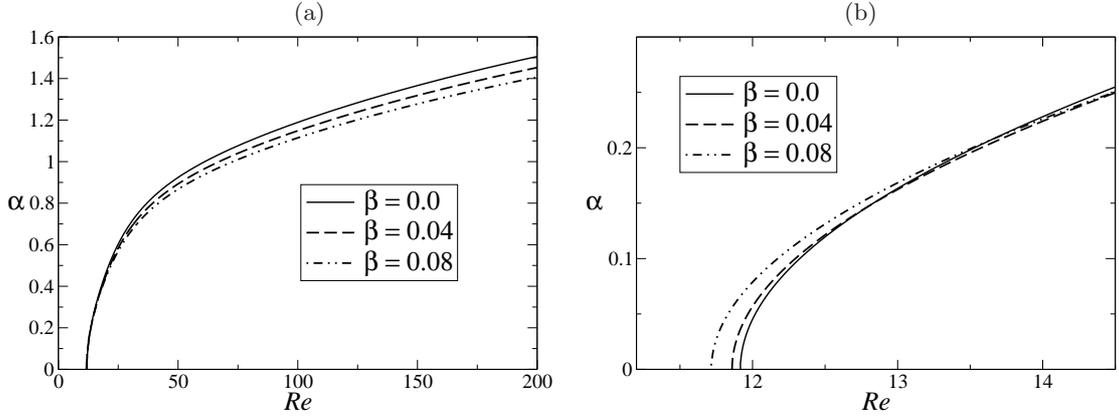

\centering
\hspace{0.2in} (a) \hspace{2.6in} (b) \\
\includegraphics [scale = 0.3]{fig3a.eps} \hspace{0.2cm}
\includegraphics [scale = 0.3]{fig3b.eps}
\caption{Neutral stability boundaries of surface mode (a) for a single fluid flow ($m = 1$) down a rigid ($\beta = 0$) and a slippery ($\beta \neq 0$) incline; (b) zoom of (a) in the range $10 \leq Re \leq 20$. The other parameters are taken as $\theta = 4^{\circ}, \, S = 0$.}
\label{fig3}
\end{figure}     

The stability results are first obtained for a single fluid film over a rigid/slippery substrate ($m = 1$) and it exactly agreed with those of Yih \cite{Yih63} ($\beta = 0$) and Samanta $et$ $al.$ \cite{Samanta11a} ($\beta \neq 0$) after taking into account the velocity scales used in the above investigations (Figs. \ref{fig3}(a),(b); $\theta = 4^{\circ}, \, S = 0$). Slip decreases the critical Reynolds number, indicating its role in destabilizing the flow system at the onset. However, beyond the threshold, slip stabilizes the flow by decreasing the range of unstable wave numbers.   

\begin{figure}[!ht]
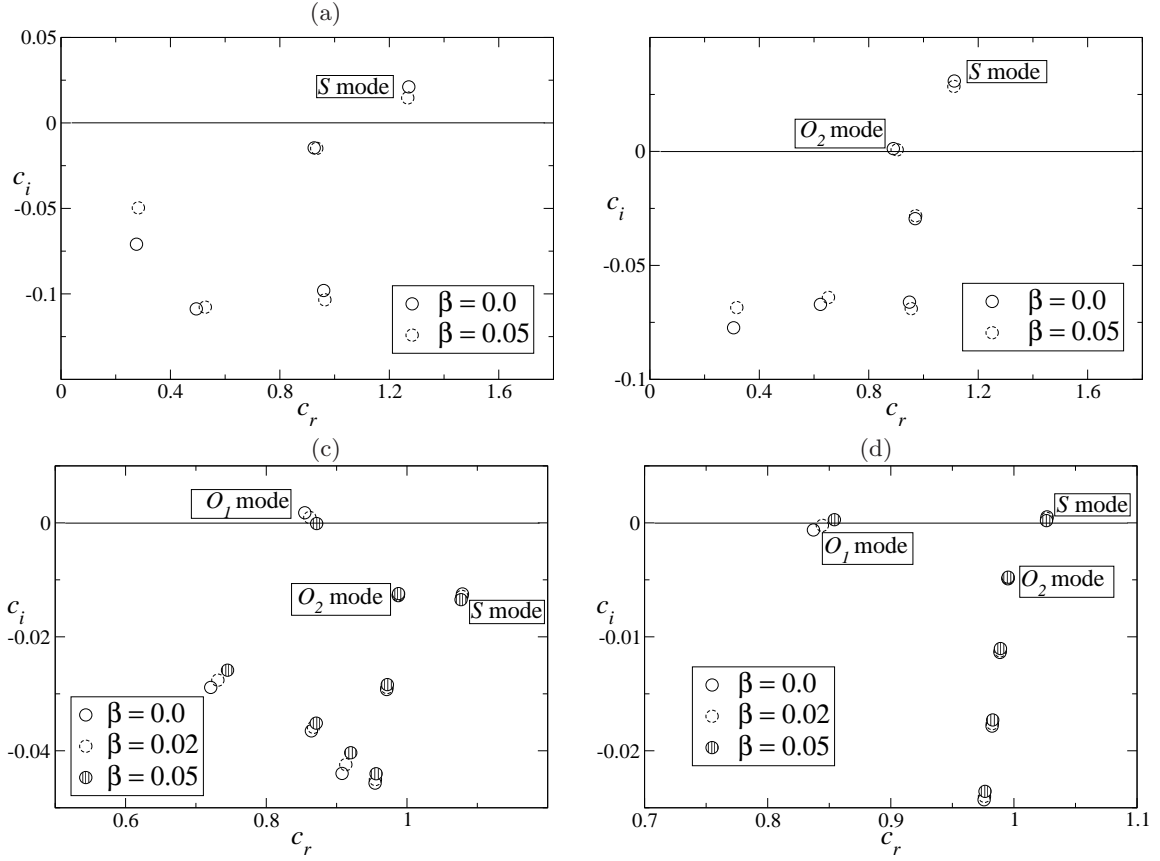

\centering
\hspace{0.2in} (a) \hspace{2.6in} (b) \\
\includegraphics [scale = 0.3]{fig4a.eps}\hspace{0.5cm}
\includegraphics [scale = 0.3]{fig4b.eps}\\
\hspace{0.2in} (c) \hspace{2.6in} (d) \\
\includegraphics [scale = 0.3]{fig4c.eps}\hspace{0.5cm}
\includegraphics [scale = 0.3]{fig4d.eps}\\
\caption{Influence of slip $\beta$ on the eigenvalues for different $Re$ and $\alpha$: (a) for $Re = 50, \, \alpha = 0.5$; (b) for $Re = 200, \, \alpha = 0.5$; (c) for $Re = 200, \, \alpha = 3.5$ and (d) for $Re = 1300,\,\alpha = 3.5$. All other parameters are $m = 1.5,\,Sc = 20,\, \theta = 10^{\circ},\,h = 0.4,\,q = 0.2$ and $S = 0$.} 
\label{fig4}
\end{figure}

The eigenspectra for $m= 1.5$ is presented in Fig. \ref{fig4}(a) - Fig. \ref{fig4}(d) for different Reynolds numbers ($Re$) and wave numbers ($\alpha$), when the mixed layer is located at $h = 0.4$ with $q = 0.2$ and $Sc = 20,\, \theta = 10^{\circ}$. The Reynolds number is based on the average viscosity and velocity across the film. At $Re = 50, \, \alpha = 0.5$, there is an unstable surface mode ($S$ mode) with phase speed $c_r > 1$ for both $\beta = 0$ and $\beta = 0.05$ (Fig. \ref{fig4}(a)). The slip at the substrate decreases the growth rate ($c_i$) of the $S-$mode at this $Re$ and $\alpha$.

As the  Reynolds number $Re$ is increased to $Re = 200$ (Fig. \ref{fig4}(b)), there occurs a second unstable mode but with $c_r < 1$, for the same wave number $\alpha = 0.5$. As the presence of this unstable mode is observed at small values of $\alpha$ but at higher $Re$, this corresponds to $O_2$ mode (in tune with notation used in Usha $et \, al.$ \cite{Usha13a}). This mode comes into existence due to the overlap of critical layer with the mixed layer. The critical layer is a thin layer around the critical point $y = y_c$ where $U_B(y = y_c) = c_r$. An increase in phase speed ($c_r$) and a decrease in growth rate ($c_i$) due to the presence of wall slip are evident. 
-

Now, with $Re$ fixed at $Re = 200$, an increase in $\alpha$ to $\alpha = 3.5$ shows the existence of another unstable overlap mode (called the $O_1$ mode) with phase speed $c_r < 1$ but different from that of $O_2$ mode. It shows the existence of short wave instability at this $Re$. An increase in slip parameter $\beta$ decreases the growth rate ($c_i$) but increases the phase speed $c_r$ of the two dimensional disturbance, is evident from Fig. \ref{fig4}(c).

With further increase in Reynolds number to $Re = 1300$ (Fig. \ref{fig4}(d)), the occurrence of three modes, namely the $S$ mode (with $c_r>1$ and $c_i>0$), the $O_2$ mode (with $c_r<1$ and $c_i<0$) and the $O_1$ mode (with $c_r<1$ and $c_i<0$ for $\beta = 0$, $c_i>0$ for $\beta = 0.05$) is observed. While the growth rate of the $S$ mode decreases with an increase in slip, the growth rate of $O_1$ and $O_2$ modes increase with an increase in slip at this large value of $Re$. 

The eigenspectra for $m = 1.5$ clearly indicate that the occurrence of the unstable modes as well as their stability characteristics are highly influenced by inertia, the presence of slip at the substrate and the range of wave numbers.

\begin{figure}[!ht]
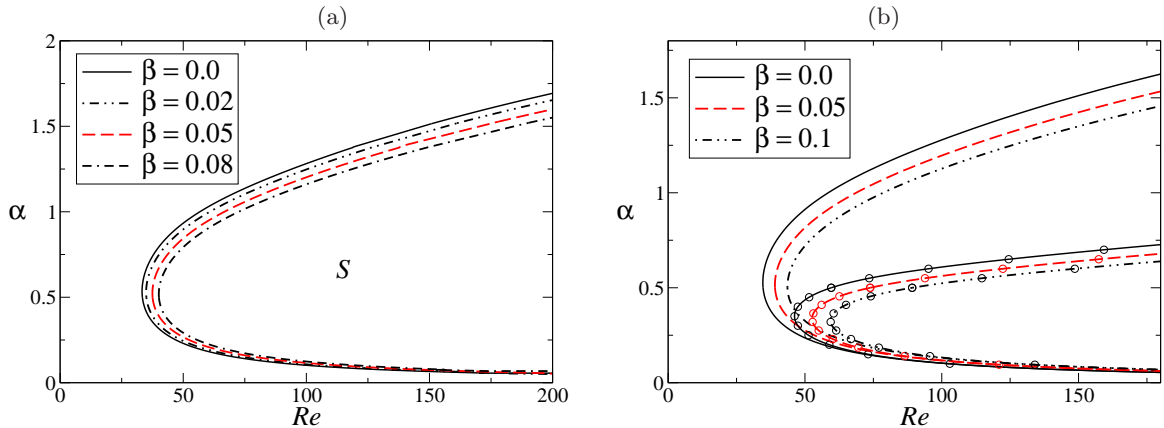

\centering
\hspace{0.2in} (a) \hspace{2.6in} (b) \\
\includegraphics [scale = 0.3]{fig5a.eps}\hspace{0.5cm}
\includegraphics [scale = 0.3]{fig5b.eps}\\
\caption{Neutral stability boundaries of the surface mode($S$ mode) for $m = 1.5,\,h = 0.4\, Sc = 20, \, q = 0.2$ and $\theta = 10^{\circ}$. (a) Effects of slip($\beta$) when $S = 0$ ; (b) $\beta$ effects when $S = 10$ (without symbols) and $S = 100$ (with symbols).}
\label{fig5}
\end{figure} 

The effects of slip on the neutral stability boundaries for the $S$ mode is presented in Fig. \ref{fig5}(a), when $m = 1.5,\,h = 0.4,\, q = 0.2,\,Sc= 20,\,\theta = 10^{\circ}$ and $S = 0$. When a more viscous fluid is adjacent to the free surface, slip stabilizes the miscible two-layer system by increasing the critical Reynolds number at the onset and decreasing the bandwidth of unstable wave numbers beyond threshold for the surface mode. There is no long-wave ($\alpha \to 0$) instability of the above system for all $\beta$, which is in contrast with the single fluid free surface flow over a rigid/slippery substrate (see Fig. \ref{fig3}(a),(b)). The influence of diffusivity on the $S$ mode instability when there is wall slip is analogous to that in the absence of slip.

\begin{figure}[!t]
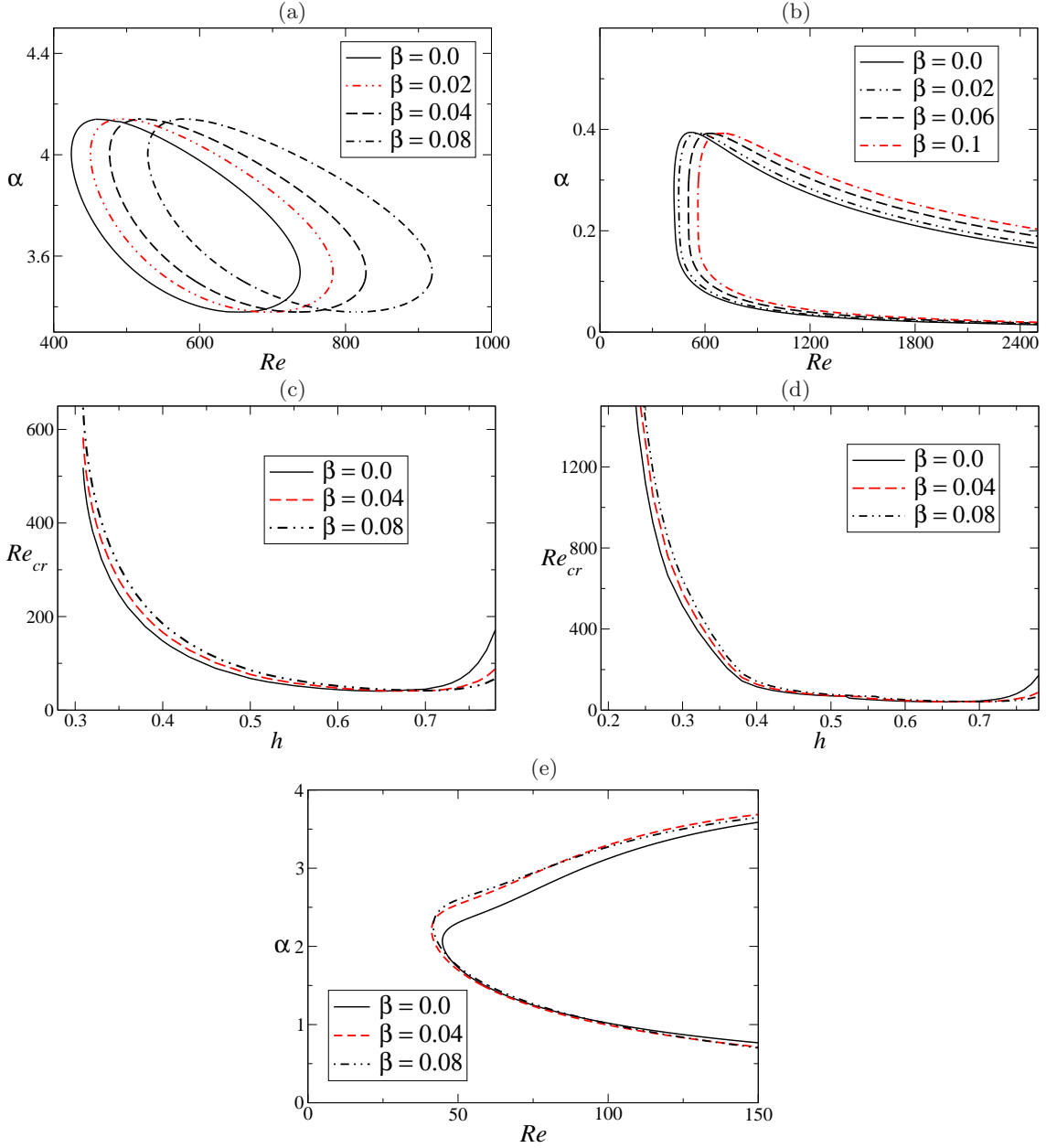

\centering
\hspace{0.2in} (a) \hspace{2.6in} (b) \\
\includegraphics [scale = 0.3]{fig6a.eps}\hspace{0.5cm}
\includegraphics [scale = 0.3]{fig6b.eps}\\
\hspace{0.2in} (c) \hspace{2.6in} (d) \\
\includegraphics [scale = 0.3]{fig6c.eps}\hspace{0.5cm}
\includegraphics [scale = 0.3]{fig6d.eps}\\
\hspace{0.2in} (e)\\
\includegraphics [scale = 0.3]{fig6e.eps}
\caption{Neutral stability maps for $h = 0.315$ with different slip parameters ($\beta$) (a) for $O_1$ mode, (b) for $O_2$ mode. Critical Reynolds number as a function of $h$ (c) for $O_1$ mode, (d) for $O_2$ mode. Fig. (e) presents the neutral stability boundaries of $O_1$ mode for $h = 0.7$. The other parameters are fixed as $m = 1.5,\,Sc= 20,\,\theta = 10^{\circ},\,q = 0.2$ and $S = 0$.}
\label{fig6}
\end{figure}

It is possible to stabilize the $S$ mode for the miscible two-fluid flow over a slippery substrate in the presence of surface tension at the free surface is clear from Fig. \ref{fig5}(b), when $m = 1.5,\,h = 0.4,\, q = 0.2,\,Sc= 10,\,\theta = 10^{\circ}$ for different values of $\beta$. The critical Reynolds number increases with an increase in $S$. At a fixed Reynolds number, there is a range $[\alpha_L, \alpha_U]$ of wave numbers in which the $S$ mode is unstable for all values of surface tension parameter ($S$) and beyond the instability threshold, $\alpha_U$ decreases significantly with an increase in $S$. Therefore, higher the surface tension, more stable is the system ($S$ mode) at moderate wave numbers. It is interesting to note that while wall slip is able to stabilize long as well as short waves at any $Re$, the presence of surface tension at the free surface is able to stabilize short waves only. The slip at the wall enhances the stabilizing role of surface tension. 

How does slip at the substrate influence the overlap modes? and Fig. \ref{fig6} provides an answer. The role of slip on the neutral stability maps of the $O_1$ mode is displayed in Fig. \ref{fig6}(a) for the configuration with $m = 1.5,\,h = 0.315,\, q = 0.2,\,Sc= 20,\,\theta = 10^{\circ}$ and $S = 0$. The critical Reynolds number ($Re_{cr}$) for the $O_1$ mode (which exist for higher wave numbers and moderate wave numbers) increases with an increase in slip parameter. This shows slip has stabilizing effect on the $O_1$ mode at the onset. But, the unstable region is shifted towards higher Reynolds numbers. As a result, the effect of slip is to destabilize the $O_1$ mode for higher Reynolds numbers. There is no significant change in the bandwidth of unstable wave numbers. 

The $O_2$ mode that exists at small wave numbers and at moderate Reynolds numbers (Fig. \ref{fig6}(b); $m = 1.5,\,h = 0.315,\, q = 0.2,\,Sc= 20,\,\theta = 10^{\circ}$ and $S = 0$) is influenced by the wall slip and it is seen that slip $(i)$ stabilizes the $O_2$ mode at the onset by increasing the critical Reynolds number, $(ii)$ stabilizes the long waves and $(iii)$ destabilizes the short waves. 

The stability properties of both $O_1$ and $O_2$ modes are affected by the location of the mixed layer ($y = h$; see Fig. \ref{fig6}(c),(d)). The parameters are the same as in Fig. \ref{fig6}(a) and (b). For each $\beta$, $Re_{cr}$ for the $O_1$ mode decreases with an increase in $h$ upto $h = h_{cr}$. This $h_{cr}$ increases increase in slip. However, for any location of the mixed layer ($y = h$) with $h < h_{cr}$, $Re_{cr}$ increases with an increase in slip, thus stabilizing the $O_1$ mode (Fig. \ref{fig6}(c)). Note that, for this set of parameters, $O_1$ mode occupies a distinct region in the $Re-\alpha$ plane and appears at higher wave numbers. For these values of $h$, $O_2$ mode is also stabilized by slip at the substrate (Fig. \ref{fig6}(d)). When the mixed layer is very close to the slippery substrate ($h > h_{cr}$), the effect of slip is reversed for both the $O_1$ and $O_2$ modes. In fact, the $O_1$ and $O_2$ modes coalesce at higher values of $h$ ($h > h_{cr}$). The effect of wall slip on the neutral stability maps for $h = 0.7$ are presented in Fig. \ref{fig6}(e). The destabilizing role of slip is observed at the onset of $O_1$ mode instability. It is also observed that critical wave numbers increase with respect to slip at higher values of $h$ (figures not shown). 

\begin{figure}[!ht]
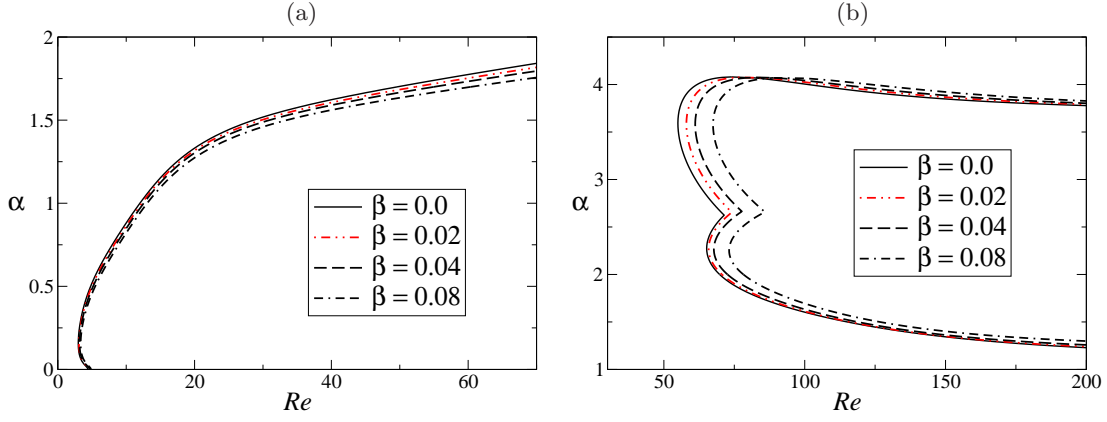

\centering
\hspace{0.2in} (a) \hspace{2.6in} (b) \\
\includegraphics [scale = 0.3]{fig7a.eps} \hspace{0.2cm}
\includegraphics [scale = 0.3]{fig7b.eps}
\caption{Effects of slip parameter ($\beta$) for antilubrication case ($m = 0.67$) when $h = 0.4,\, q = 0.2,\,\theta = 10^{\circ}, \, Sc = 20$ and $S = 0$: (a) on the surface modes and (b) on the overlap modes.}
\label{fig7}
\end{figure}  

We now focus on the antilubrication configuration ($m<1$) in which more viscous fluid is adjacent to the slippery substrate. There is an inflectional point for each $\beta$ (both $\beta = 0$ and $\beta \neq 0$) and the base velocity is concave in the mixed layer region (Fig. \ref{fig2}(a)). At any $y$ within the film $U_B(y)$ for $m < 1$ is smaller than that for the lubrication configuration ($m>1$). However, slip increases the wall velocity as a result, wall stress decreases with slip.

We perform computations with $m<1$ ($m = 0.67$) and phase speed greater than the base velocity at the free surface to check for the existence of surface mode ($S$ mode) for this configuration. Fig. \ref{fig7}(a) shows the neutral stability boundaries of the $S$ mode for $m = 0.67$ with different values of slip parameter $\beta$. The other parameters are fixed at $h = 0.4,\, q = 0.2,\,\theta = 10^{\circ}, \, Sc = 20$ and $S = 0$. Beyond the threshold for instability, for any $\beta$, both the long and short waves are destabilized for the system with $m = 0.67$. This is in contrast to the case $ m>1$ ($m = 1.5$), where the long waves are stable for all $Re$. Further, the slip increases the critical Reynolds number $Re_{cr}$ at the onset and decreases the range of unstable wave numbers beyond the threshold for instability indicating the stabilizing role of slip for the case $m <1$. We also note from Fig. \ref{fig5}(a) and Fig. \ref{fig7}(a) that the critical Reynolds number $Re_{cr}$ for $m<1$ is smaller than the $Re_{cr}$ for $m>1$, for each $\beta$. Also, the range of unstable wave numbers for $m < 1$ is $[0,\alpha_c]$ while that for $m>1$ is $[\alpha_l, \alpha_u]$ with $\alpha_l > 0$ and $\alpha_c > \alpha_u$. This indicates that $m<1$ configuration is more unstable than $m>1$ configuration for the considered parameter values.                                         

The computation performed with $m<1$ ($m = 0.67$) reveals the existence of an unstable overlap mode ($O$ mode) when phase speed is smaller than the free surface base velocity. The neutral stability maps of this $O$ mode are plotted in Fig. \ref{fig7}(b) for different values of slip parameter $\beta$.  Fig. \ref{fig7}(b) shows the coalescence of $O_1$ and $O_2$ modes enclosing a big unstable region due to the total viscosity stratification under overlap condition. Slip at the substrate stabilizes the $O$ mode at the onset of the instability.

It is worth mentioning that the above results have relevance in the understanding of rupture dynamics of pre corneal tear film after a blink for a range of values of viscosity stratification parameter ($m$). The human tear film is described as comprising three distinct layers: an outermost lipid layer \cite{Berger74,Braun12}, an innermost mucus layer and an intermediate aqueous layer. The aqueous component of the tear film fills the sachs under the lower and upper lids \cite{Mishima65}. The investigations in \cite{Bron04,Cher07,Zhang03b} claim that the mucus layer blends with the aqueous layer without any interfacial tension between two. This suggests that the three layer theory can be replaced by one in which the mucins are distributed throughout the mucoaqueous layer which forms the bulk of the tear film and the epithelial mucins form a complex barrier at the corneal surface. The present study can be thought of as modeling the above scenario in the following way: the epithelial mucins forming a complex barrier at the corneal surface is modeled by a substrate with velocity slip and that the mocoaqueous layer forming the bulk of tear film as a film with viscosity varying continuously within the film as described by Eq. (\ref{bs3}). The typical aqueous layer viscosity is $10^{-2} g\,cm^{-1}\,s^{-1}$ and the viscosity of the mucus layer \cite{Zhang03a} varies from $10^{-1} g\,cm^{-1}\,s^{-1}$ to $10^{-2} g\,cm^{-1}\,s^{-1}$ and this yields the viscosity ratio $m = \frac{\mu_2}{\mu_1}$ (see Fig. 1) to vary from $0.1$ to $1.0$. Typical value of $m = 0.67$ is considered in the above computation and the results reveal the existence of overlap instability due to the presence of mixed layer (see Fig. 7(b)) which can be suppress/enhance by wall slip. This instability may be regarded as leading to rupture of the tear film. The computations are also performed for values of viscosity ratio $m>1$ which are relevant in other application such as surface coating, multi-layer photographic films.
        
\begin{figure}[!t]
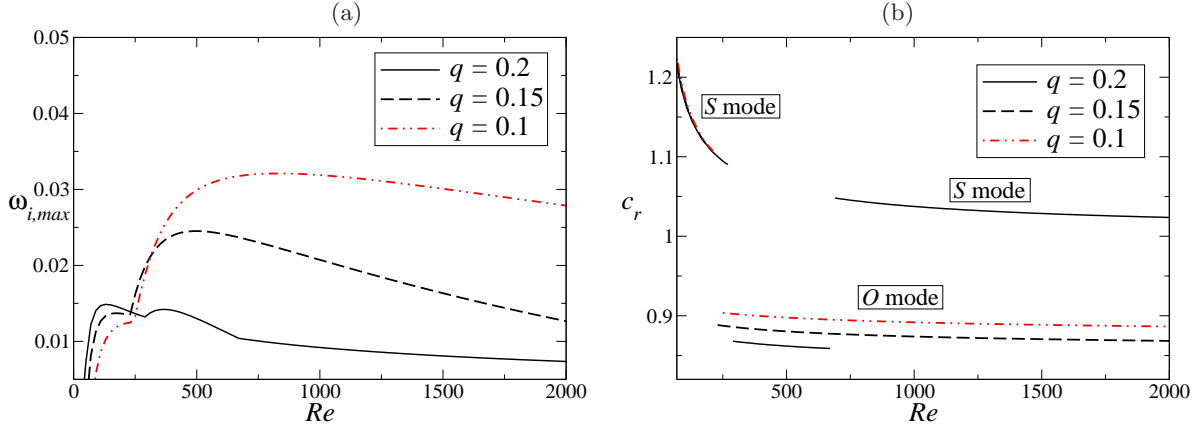

\centering
\hspace{0.2in} (a) \hspace{2.6in} (b) \\
\includegraphics [scale = 0.3]{fig8a.eps} \hspace{0.2cm}
\includegraphics [scale = 0.3]{fig8b.eps}
\caption{(a) maximum growth rate ($\omega_{i,max}$) and (b) phase speed ($c_r$), as a function of Reynolds number for  $m = 1.5,\,\beta = 0.05,\,h = 0.4,\, q = 0.2,\,Sc= 20,\,\theta = 10^{\circ}$ and $S = 0$.}
\label{fig8}
\end{figure}

In the above discussions, the mixed layer thickness ($q$) has been taken as $q = 0.2$. It is important to examine the influence of mixed layer thickness on the stability properties and Fig. \ref{fig8}(a) shows the maximum growth rate ($\omega_{i,max}$) over all wave numbers for the dominant mode as a function of Reynolds number ($Re$) with different values of $q$. The other parameters are taken as $m = 1.5,\,\beta = 0.05,\,h = 0.4,\, q = 0.2,\,Sc= 20,\,\theta = 10^{\circ}$ and $S = 0$. Note that $\omega_{i,max}>0$ implies instability of the flow system.

\begin{figure}[!t]
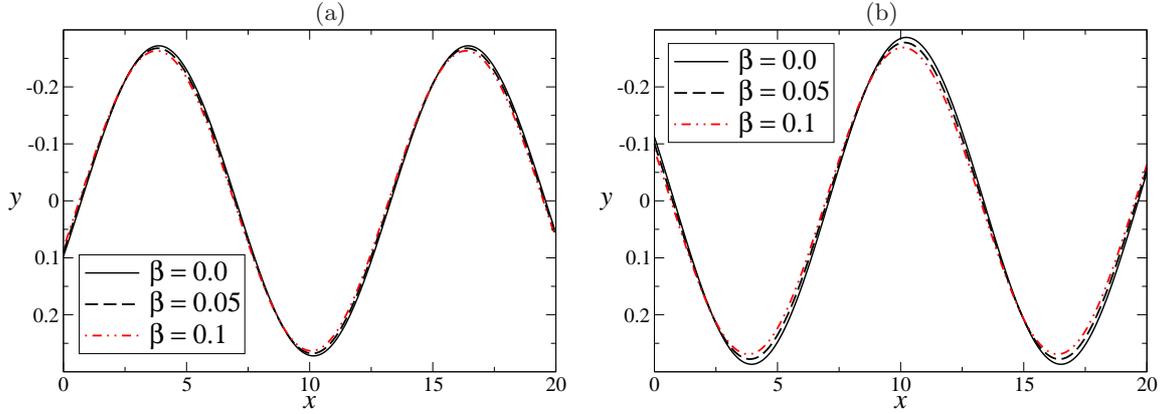

\centering
\hspace{0.2in} (a) \hspace{2.6in} (b) \\
\includegraphics [scale = 0.3]{fig9a.eps} \hspace{0.2cm}
\includegraphics [scale = 0.3]{fig9b.eps}
\caption{Effects of slip ($\beta$) on the amplitude free surface perturbation for $m = 1.5,\,h = 0.4,\, q = 0.2,\,Sc= 20,\,\theta = 10^{\circ}$ and $S = 0$. (a) At time $t = 0$; (b) at time $t = 5$.}
\label{fig9}
\end{figure}

\begin{figure}[!ht]
\vspace{0.7cm}
\centering
\hspace{0.05in} (a) \hspace{2.8in} (b)\\ 
\includegraphics [scale = 0.27]{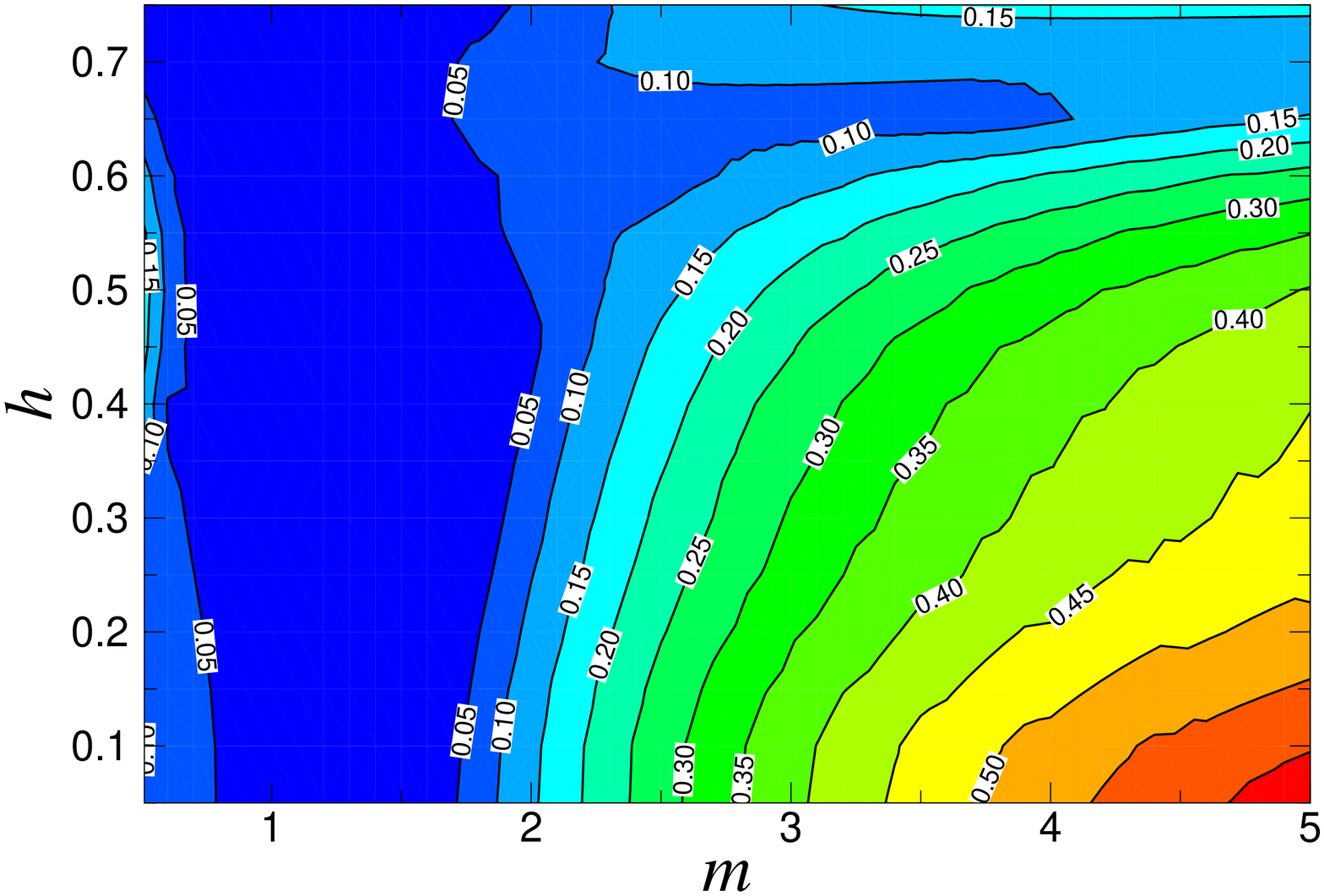}\hspace{-0.5cm}
\includegraphics [scale = 0.27]{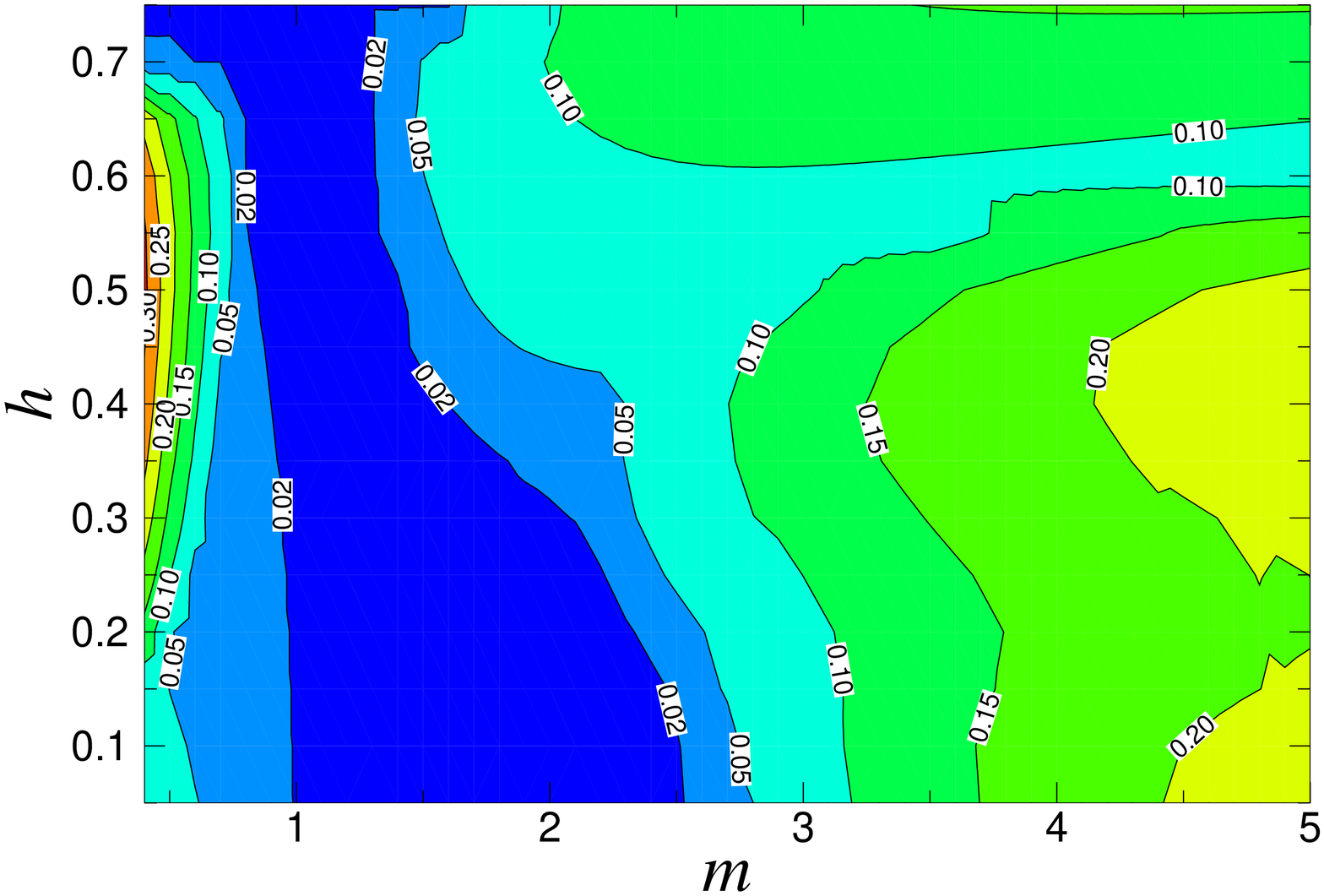}\\
\hspace{0.05in} (c) \hspace{2.8in} (d) 
\includegraphics [scale = 0.27]{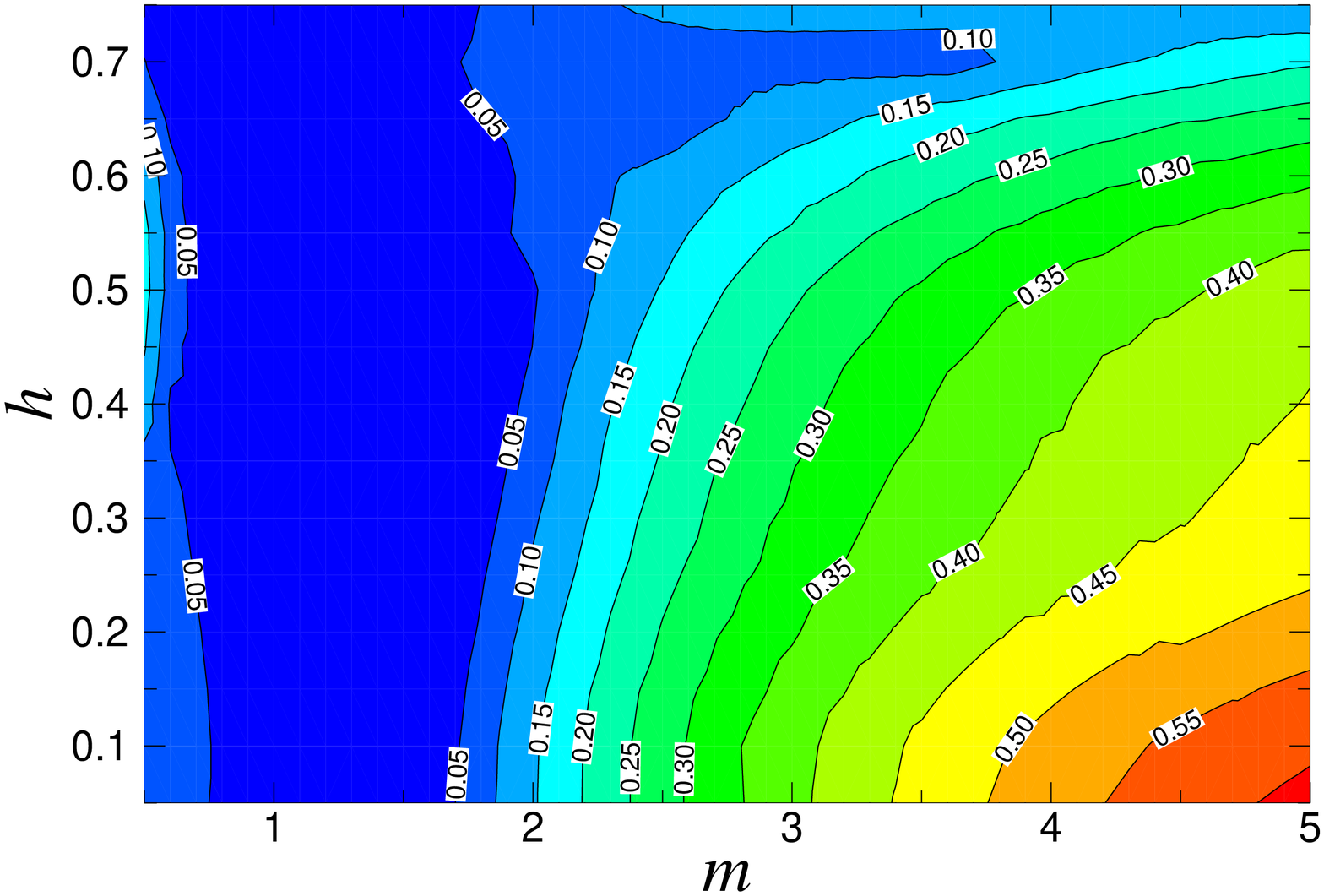}\hspace{-0.5cm}
\includegraphics [scale = 0.27]{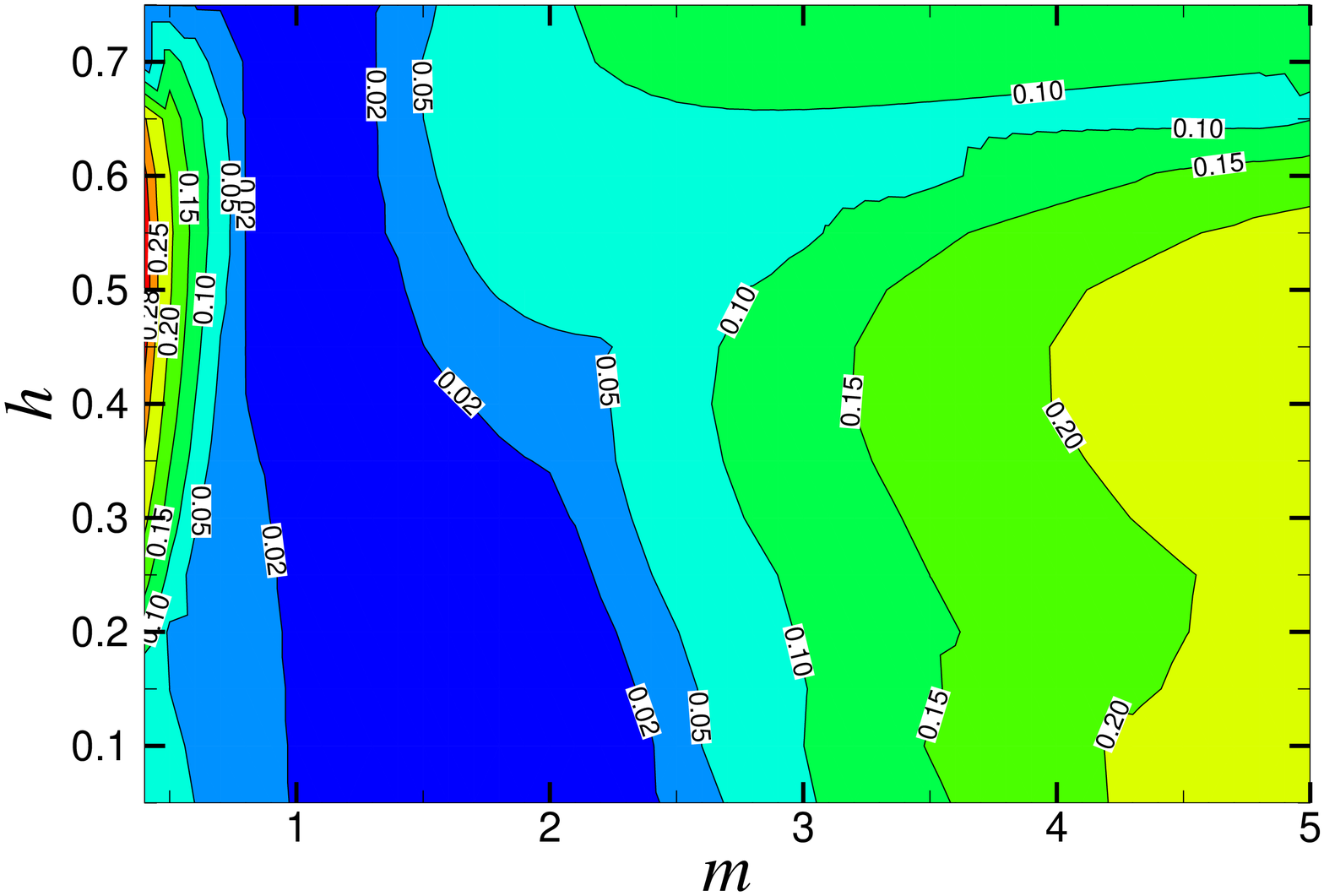}\\
\caption{Contour plots of $\omega_{i,max}$ in $m-h$ plane for two different $Re$ and $\beta$: (a) $Re = 50, \, \beta = 0.0$; (b) $Re = 200, \, \beta = 0.0$; (c) $Re = 50, \, \beta = 0.1$ and (d) $Re = 200, \, \beta = 0.1$. The other parameters are $q = 0.2,\,Sc= 20,\,\theta = 10^{\circ}$ and $S = 0$.}
\label{fig10}
\end{figure}

For a typical value of $q\, (= 0.2)$ considered here, while the $S$ mode is dominant for smaller and higher values of $Re$ (see Fig. \ref{fig8}(b); solid line with phase speed $c_r>1$), the $O$ mode is dominant for moderate Reynolds numbers (Fig. \ref{fig8}(b); solid line with phase speed $c_r<1$). As $q$ decreases, the $S$ mode is stabilizing due to the decrease in $\omega_{i,max}$ but the $O$ mode is destabilizing due to the increase in $\omega_{i,max}$ (see Fig. \ref{fig8}(a)). The phase speed for the $S$ mode is not significantly affected by changing the $q$ value but that for the $O$ mode, the phase speed increases. 

Fig. \ref{fig9}(a) and Fig. \ref{fig9}(b) display the amplitude of the free surface disturbance at time $t = 0$ (Fig. \ref{fig9}(a)) and at time $t = 5$ (Fig. \ref{fig9}(b)) when $m = 1.5,\,h = 0.4,\, q = 0.2,\,Sc= 20,\,\theta = 10^{\circ}$ and $S = 0$. The amplitude of free surface is reduced as slip increases at each $t$ and the reduction in amplitude of the free surface is enhanced by wall slip as time progress (Fig. \ref{fig9}(b); time $t = 5$).  

Fig. \ref{fig10} presents the contours of maximum growth rate $\omega_{i,max}$ in $m-h$ plane, including all modes over all wave numbers, at two $Re$ (Figs. \ref{fig10}(a),(c) with $Re = 50$ and Figs. \ref{fig10}(b),(d) with $Re = 200$) for $\beta = 0.0$ (Figs. \ref{fig10}(a),(b)) and $\beta = 0.05$ (Figs. \ref{fig10}(c),(d)). The other parameters are $q = 0.2,\,Sc= 20,\,\theta = 10^{\circ}$ and $S = 0$. Fig. \ref{fig10}(a) shows that as $m$ increases/decreases from $m = 1$, the maximum growth rate $\omega_{i,max}$ increases for any location of mixed layer. There is a range of $m$ close to $m = 1.5$ for all positions of mixed layer $h$, where the flow system is less unstable. $\omega_{i,max}$ is higher for the lubricated case with more viscous fluid adjacent to the free surface (large values of $m$) and mixed layer is located close to the free surface. The stabilizing role of slip $\beta$ at this $Re$ is also clearly seen (Fig. \ref{fig10}(c)). As $Re$ increases to $Re=200$,  $\omega_{i,max}$ does not change qualitatively (Fig. \ref{fig10}(b),(d)). The results reveal that the stability characteristics are highly influenced by the viscosity ratio $m$ and the location of mixed layer ($h$) from the free surface. Further, the slip velocity at the wall has dual role on the stability properties of miscible two fluid free surface flow and this depends on $Re,\,h$ and $m$.

The above results clearly revel the influence of a weakly space-dependent viscosity on the stability properties of the flow system and demonstrate how the instabilities that arise due to viscosity stratification can be suppressed by velocity slip at the wall. 
\begin{figure}[!b]
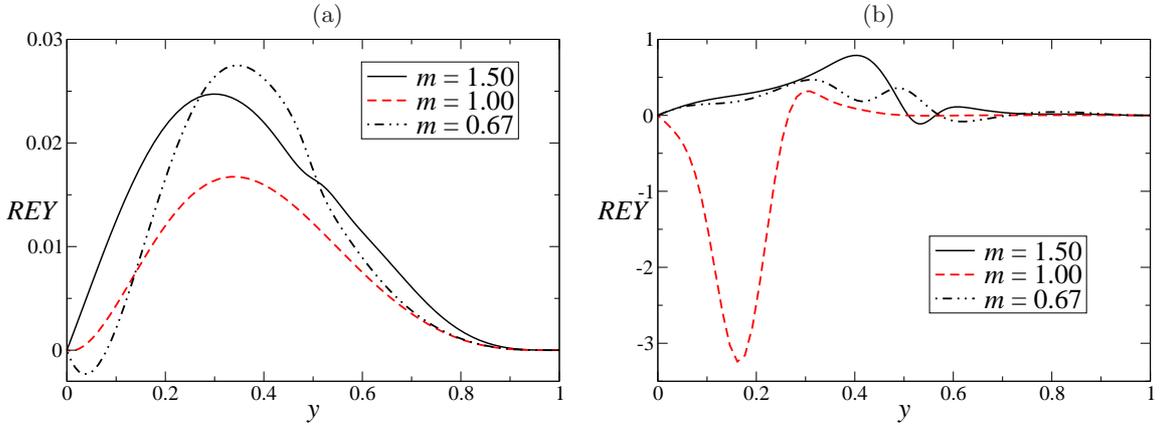

\centering
\hspace{0.2in} (a) \hspace{2.6in} (b) \\
\includegraphics [scale = 0.3]{fig11a.eps} \hspace{0.2cm}
\includegraphics [scale = 0.3]{fig11b.eps}\\
\caption{Reynolds stress ($REY$) as a function of $y$ for different viscosity ratio when $h = 0.4, \, q = 0.2,\,Sc = Pe/Re = 20$ $\beta = 0.0$ and $\theta = 10^{\circ}$. Fig. (a) for surface mode when $Re = 50$ and $\alpha = 0.5$. Fig. (b) for overlap mode when $Re = 250$ and $\alpha = 3.0$.}
\label{fig11}
\end{figure} 

\begin{figure}[!ht]
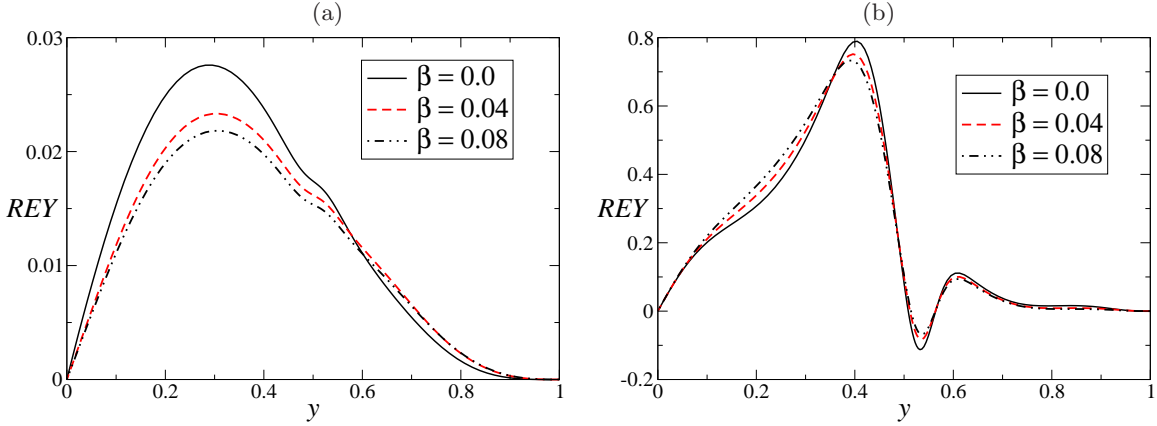

\centering
\hspace{0.2in} (a) \hspace{2.6in} (b) \\
\includegraphics [scale = 0.3]{fig12a.eps} \hspace{0.2cm}
\includegraphics [scale = 0.3]{fig12b.eps}\\
\caption{Influence of slip $\beta$ on the Reynolds stress ($REY(y)$) as a function of $y$ for $h = 0.4, \, q = 0.2,\,Sc = Pe/Re = 20$ $m = 1.5$ and $\theta = 10^{\circ}$. Fig. (a) for surface mode when $Re = 50$ and $\alpha = 0.5$. Fig. (b) for overlap mode when $Re = 250$ and $\alpha = 3.0$.}
\label{fig12}
\end{figure}

\begin{figure}[ht!]
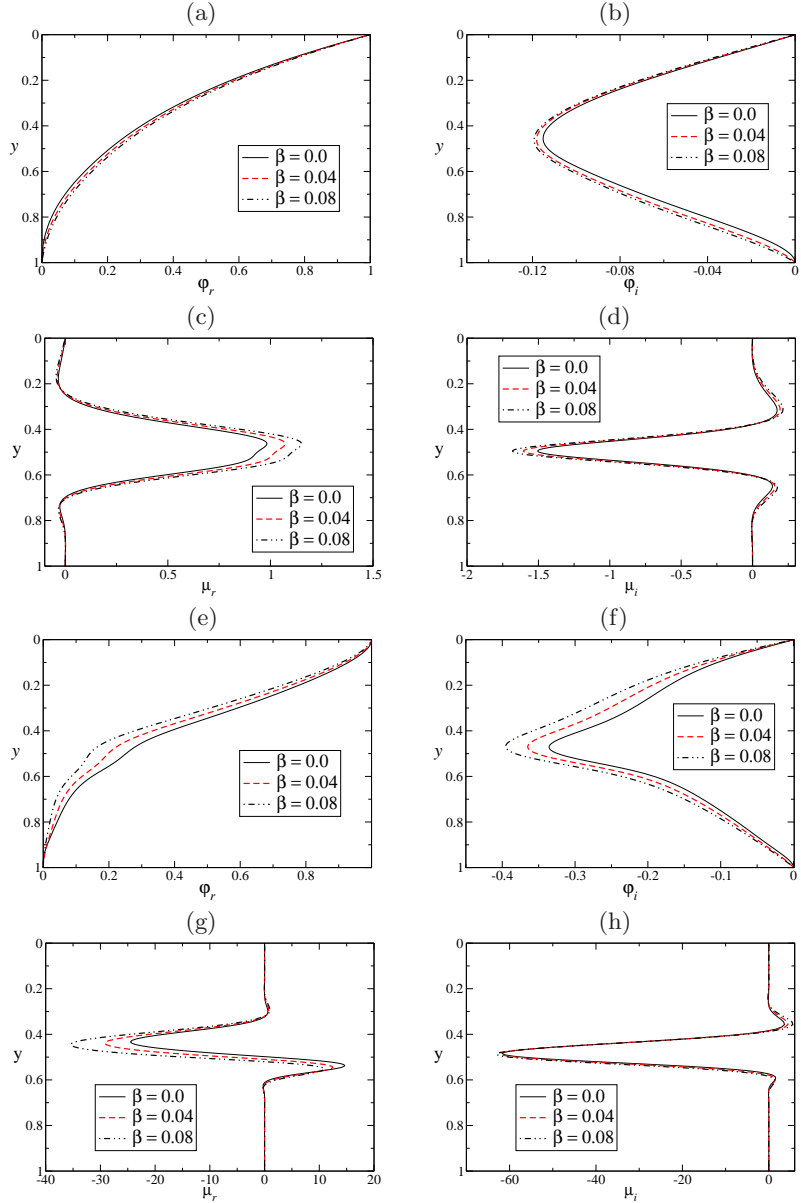

\centering
\hspace{-0.0cm} (a) \hspace{4.8cm} (b)\\
\includegraphics [scale = 0.2]{fig13a.eps}\hspace{0.69cm}
\includegraphics [scale = 0.2]{fig13b.eps}\\
\hspace{-0.0cm} (c) \hspace{4.8cm} (d)\\
\includegraphics [scale = 0.2]{fig14a.eps}\hspace{0.57cm}
\includegraphics [scale = 0.2]{fig14b.eps}\\
\hspace{-0.0cm} (e) \hspace{4.8cm} (f)\\
\includegraphics [scale = 0.2]{fig13c.eps}\hspace{0.69cm}
\includegraphics [scale = 0.2]{fig13d.eps}\\
\hspace{-0.0cm} (g) \hspace{4.8cm} (h)\\
\includegraphics [scale = 0.2]{fig14c.eps}\hspace{0.57cm}
\includegraphics [scale = 0.2]{fig14d.eps}\\
\caption{Effects of $\beta$ on the real and imaginary parts of amplitude of the stream function ($\phi_r$, $\phi_i$) and the viscosity perturbation ($\mu_r$, $\mu_i$) for $h = 0.4, \, q = 0.2,\,m = 1.5$ $Sc = Pe/Re = 20$ and $\theta = 10$. Figs. (a)-(d) for surface mode when $Re = 50$ and $\alpha = 0.5$. Figs. (e)-(h) for overlap mode when $Re = 250$ and $\alpha = 3.0$.}
\label{fig13}
\end{figure}  

At this stage, we consider the behaviour of the Reynolds stress $REY(y)$ (production energy term) \cite{Malik05a,drazinandreid} given by,
\begin{equation}
REY(y) = -\frac{\mathrm{i}\alpha}{4}\, U_B{'}({\bar{\phi}\phi^{'}} - \phi \bar{\phi^{'}}),
\nonumber
\end{equation}
and the details are presented in Fig. \ref{fig11} and Fig. \ref{fig12} as a function of $y$ for the surface mode and the overlap mode. The corresponding eigenfunctions for the stream function perturbation and viscosity perturbation are presented in Fig. \ref{fig13} when  $h = 0.4, \, q = 0.2,\,Sc = Pe/Re = 20$, $\beta = 0.0$ and $\theta = 10^{\circ}$. The Reynolds stress, $REY(y)$ is positive within the film and this confirms the existence of surface mode (Fig. \ref{fig11}(a)) instability in the free surface flows ($m = 1$ or $m \neq 1$). This production term is influenced by viscosity contrast away from the wall and it is significant in the mixed layer ($0.4\leq y \leq 0.6$). While the surface mode exists for both the single fluid ($m = 1$) and the miscible two-fluid ($m \neq 1$) free surface flow, the $O_1$ mode exists only for miscible two-fluid free surface flow for high wave numbers. This is reflected in Fig. \ref{fig11}(b) for values of $m \neq 1$. The above observation is for free surface flows over a rigid substrate ($\beta = 0$). Fig. \ref{fig12}(a) and (b) present the effect of slip ($\beta$) on the production term ($REY(y)$) when more viscous fluid is close to the free surface ($m = 1.5$) for the surface and $O_1$ modes respectively. It is clear that the production energy term is not affected near the wall by increase in slip. In view of this, one may think that slip has no influence on the stability characteristics of single/miscible two-fluid free surface flows. However, we observed influence of slip away from the wall.

The above results are presented for a fixed Reynolds number $Re$ ($Re = 50$ for $S$ mode and $Re = 250$ for $O_1$ mode) and wave number $\alpha$ ($\alpha = 0.5$ for $S$ mode and $\alpha =3.0$ for $O_1$ mode). We have seen from above results that slip has influence on the production energy term away from the wall. What is the cause for this decrease in $REY$ needs to be understood. For a fixed Reynolds number ($Re$), when slip ($\beta$) increases, the base flow accelerates and film thickness decreases (since $Re = \frac{\rho U H}{\mu_1}$ where $\rho, \, \mu_1$ are fixed) resulting in overall production energy which in turn acuses a reduction in $REY(y)$.

\section{Conclusions}
\label{conl}
The role of wall slip on the stability of miscible two-fluid flow down a slippery inclined substrate is examined. Slip at the wall shows a promise for stabilizing the surface mode of the flow system for any viscosity ratio ($m$). This is in contrast with the single fluid flow down a slippery substrate where wall slip has a destabilizing role at the onset of the instability \cite{Samanta11a}. 

The existence of overlap modes ($O_1/O_2$ mode for large/small wave numbers) under overlap condition, as in the rigid substrate case is shown; however, there is delay in the occurrence of these modes in the presence of wall slip when the mixed layer is not very close to the slippery wall. Slip pushes the boundary of the unstable region for the $O_1$ mode towards higher Reynolds numbers for the lubrication case ($m>1$). Destabilizing effect of slip on the $O_1$ mode has been found when the mixed layer is very close to the slippery substrate ($h \geq 0.7$); for these values of $h$, the two overlap modes coalesce and becomes the dominant mode. The stabilizing or destabilizing role of slip exhibited in the results depends on the location of the mixed layer, the viscosity ratio and the diffusivity parameter.

The stabilizing effect of surface tension is enhanced by the presence of wall slip by delaying the onset of instability and reducing the range of unstable wave numbers. The slip effects are not significant at higher Schmidt numbers.

The production energy term ($REY(y)$) as a function of $y$ has its maximum closer to the free surface rather then near the slippery substrate in contrast to that in the case of miscible two-fluid Poiseuille flow in a channel where the production term is maximum near the wall (see in Govindarajan $et$ $al.$ \cite{Govindarajany03}). This may be the cause for weak/mild effect of slip for free surface flows of miscible two-fluid flow (and free surface single fluid flow).   
\section*{Acknowledgement}
The authors sincerely thanks to Prof. Rama Govindarajan (TIFR Centre for Interdisciplinary Sciences, India) for providing the base numerical code and many useful discussions.


\end{document}